# Extending UML-RT for Control System Modeling

Qimin Gao, L.J. Brown and L.F. Capretz
Department of Electrical and Computer Engineering, The University of Western Ontario
London, Ontario, Canada N6A 5B9

**Abstract:** There is a growing interest in adopting object technologies for the development of real-time control systems. Several commercial tools, currently available, provide object-oriented modeling and design support for real-time control systems. While these products provide many useful facilities, such as visualization tools and automatic code generation, they are all weak in addressing the central characteristic of real-time control systems design, i.e., providing support for a designer to reason about timeliness properties. We believe an approach that integrates the advancements in both object modeling and design methods and real-time scheduling theory is the key to successful use of object technology for real-time software. Surprisingly several past approaches to integrate the two either restrict the object models, or do not allow sophisticated schedulability analysis techniques. This study shows how schedulability analysis can be integrated with UML for Real-Time (UML-RT) to deal with timing properties in real time control systems. More specifically, we develop the schedulability and feasibility analysis modeling for the external messages that may suffer release jitter due to being dispatched by a tick driven scheduler in real-time control system and we also develop the scheduliablity modeling for sporadic activities, where messages arrive sporadically then execute periodically for some bounded time. This method can be used to cope with timing constraints in realistic and complex real-time control systems. Using this method, a designer can quickly evaluate the impact of various implementation decisions on schedulability. In conjunction with automatic code-generation, we believe that this will greatly streamline the design and development of real-time control systems software.

## INTRODUCTION

Real-time control systems are concurrent systems with timing constraints. They have widespread use in industrial, commercial and military applications. They require both logical correctness and timing correctness, the logical correctness can be expressed in terms of correct input and output, the timing correctness can be expressed that the system must meet the time-critical deadlines to prevent a catastrophic system failure. Real-time control systems are one kind of hard real-time systems. They are differentiated from other types of systems by the timing requirements associated with some or all of their computations. As a result, validating such systems requires that these additional timing constraints also be satisfied. This verification is especially necessary for the real-time control systems, where fatal situations may occurs if any timing constraints are not met. Typically, the designer of real-time control systems has dealt with these timeliness properties by using their intuitive engineering skills to design such systems and then by substantiating their design through systems simulation. While this method produces the desired effect after possibly several iterations, it greatly relies on the abilities of the designer and unnecessarily consumes an elaborate amount of time and effort.

To eliminate these shortcomings, there have been many attempts to make use of object-oriented technology for real-time software. Some of them have come from the industrial area [1-4], while others have come from academia [5-7]. Many of these claims are mostly based on assumption that real-time scheduling theory can be used to perform schedulability analysis. But, traditional real-time scheduling theory results can be directly used only when the object models are restricted to look like the tasking models employed in real-time scheduling theory, as it has been done in [6]. In other cases, either the claims are unsupported [3] or based on less sophisticated analysis [5]. The first attempt to apply real-time scheduling theory to object-oriented design using the state-of the art in the both fields is described in [8], which shows how to integrate traditional scheduliability analysis techniques with object-oriented design models based on the assumptions that the entire external message arrives perfectly on periodic or aperiodic time interval. The software presented in [9] implements scheduling theory for UML model design by using the technologies described in reference [8], these integrated tools allow issues on timeliness to be addressed much earlier on in the development process. However, some critical issues regarding real-time control systems are not well addressed by the current approaches, especially because schedulability analysis for real-time control systems has not been effectively incorporated. Although some researchers [8, 9] have addressed this problems by providing code synthesis of scheduling aspects and functionality aspects models, they have mainly focused on the assumptions that all external events arrives perfectly on periodic or aperiodic without release jitter and sporadic effects. In general the real–time control systems are not the case, a message may be delayed by the polling of a tick scheduler, or perhaps awaiting the arrival of a message and some real-time control systems have messages that behave as so-called sporadically periodic; a message arrival at some time, executes periodically for a bounded number of periods and then re-arrives periodically for a number of times and then does not re-arrive for a larger time. Examples of such messages are interrupt handlers for burst interrupts or certain monitoring messages in real-time control systems. Until now there is no extended method of the object-oriented design methodologies to deal with these timing constraints for real-time control systems. Thus the above analysis methods need to be improved.

In this study, we will present an approach to incorporating schedulability analysis in a UML for Real-Time (UML-RT) model-based development process [10]. Using this approach, satisfaction of the end-to-end timing constraints of real-time control systems can be verified and the schedulability analysis results will be used for aspect-oriented code generation in the model transformation and automatic code generation.

**UML-RT and Real-Time Scheduling Theory:** The Unified Modeling Language (UML) is a graphical modeling language for visualizing, specifying, constructing and documenting the artifacts of software systems. The UML is a widely accepted language and it is becoming a de-facto standard for object-oriented modeling. UML has a strong set of general purpose modeling language concepts and has been designed as an open-ended language applicable across different domains. The tool, named UML-RT for real-time, developed by the Rational Corporation, uses UML to express the original ROOM (Real-Time Object-Oriented Modeling) concepts and their extensions.

UML-RT uses the notion of capsules to describe concurrent, active objects. Capsules are objects that communicate with other capsules through interfaces called ports and have each their own thread of execution. Capsules differ from other classes in that capsules can call operations on classes. Sending messages through public ports is the only method that capsules can communicate with other capsules. In addition to that, capsules have their behavior defined by UML hierarchical state machines (whereas classes have their behavior defined by methods). The collaborative behavior of the collection of sub-capsules can be described in a number of ways. Sequence diagrams illustrate capsule interactions through message exchanges in a time sequence. Every capsule in the sequence diagram has a lifeline. Time progresses from top to bottom along a lifeline. Sequence diagrams use directed message arrows to describe messages sent from one capsule to another. The horizontal dimension represents the different objects in the interaction.

Scheduling theory for real-time systems has received a great deal of attention. The first contribution to real-time scheduling theory was made by [11]. They developed optimal static and dynamic priority scheduling algorithm for hard real-time systems. In general, real-time scheduling theory models are centered on end-to-end system behavior, which are modeled using the notion of tasks. A task represents an entity requiring execution in some specified environment and it has several characteristics affiliated with it. Basically, scheduling theory modeling expresses a real-time control system as a collection of tasks. Since then, significant progress has been made on generalizing and improving the schedulability analysis. The authors developed exact schedulability analysis to determine worst-case timing behavior for tasks with hard real-time constraints in the RMA model considered in their initial work [11], as well as extended models, such as arbitrary deadlines, release jitter, sporadic and periodic tasks [12-18].

Most of the deterministic schedulability analysis techniques follow the same approach. First, the notion of the critical instant of a task is defined to be an instant at which a request for that task will have the largest response time. Then, the notion of busy period at level ‘ $i$ ’ is defined to be a continuous interval of time during which events of priority ‘ $i$ ’ or higher are being processed [11]. With these concepts, the calculation of the worst-case response time of an action involves the computation of the response time for successive arrivals of the action, starting from a critical instant until the end of the busy period, also the response time of a particular instant of action can be calculated by considering the effects of the blocking factor from lower priority actions and the interference factor from higher or equal priority actions, including the previous instance of the same action. If the worst-case response time of the action is less than or equal to its deadline, the action can be said to be schedulable and feasible. Otherwise, the action is not schedulable or feasible.

**An Example of Control System:** For instance, Fig. 1 depicts a typical reverse rolling mill in the steel rolling mill. It has a payoff reel, a rolling mill and a tension reel. A hot coil strip is uncoiled by the payoff reel. The strip is rolled to the specified thickness and coiled by the tension reel. The aim of the rolling process is to reduce the thickness of a strip to a desired thickness gauge. This is done by applying a force to the strip while moving through the roll gap. In order to meet increasing demand for the high precision of strip thickness, a new automatic gauge control system was developed with containing *Roll Gap Control, Roll Speed Control and Roll Eccentricity Compensation.* The *Roll Gap Control System* attempts to adjust the force from the hydraulic cylinder and hence the roll gap, to ensure the output thickness of the rolled strip. The *Roll Speed Control System* automatically adjusts the roll speed according to the mass flow theory and the tension of the steel strip to reduce the influence of thickness fluctuation and satisfy the high quality requirements. The roll *eccentricity compensation system* is applied to adjust the roll gap according the right compensation amplitude. If the eccentricity compensation is not done as the right value, it cannot cancel the effect of eccentricity in the rolling process; it can make the strip thickness become worse. The eccentricity compensation must be done in the right time or right phase. Even if it is done in the right amplitude, but it is not done at the right time, it can also make the strip thickness worse. All the control systems must guarantee their functional requirements and timing requirements. In order to design such systems, we will use the object–oriented analysis and design methodologies to analysis the functional requirements and timing requirements in such real-time control systems.

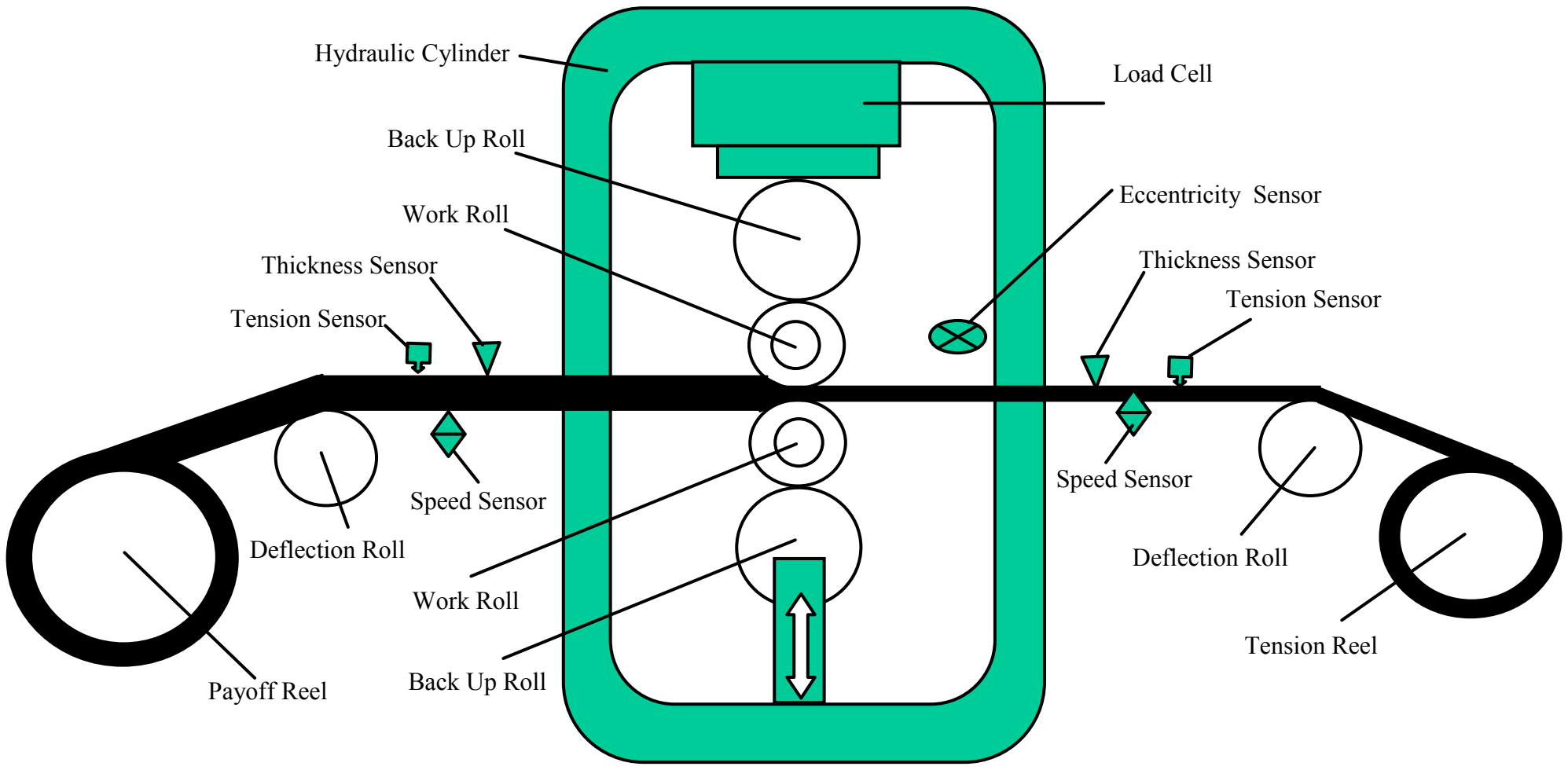

Fig. 1: The Reverse Rolling Mill

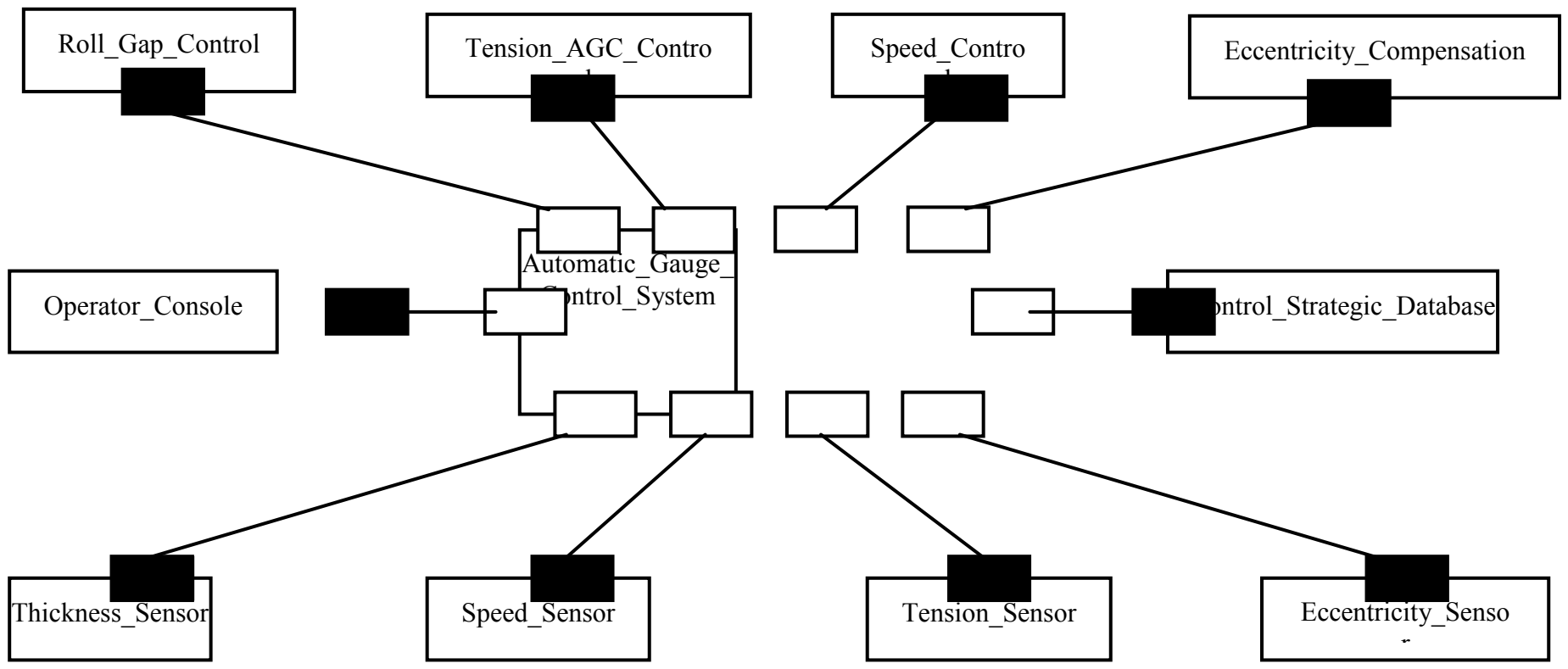

Fig. 2: Structure Model of Automatic Gauge Control System

**Control System Modeling in UML-RT:** The basic architecture entity in UML-RT is an active object; these active objects are called capsules. These capsules interact with each other only through sending and receiving messages via interface objects called *ports* that are specialized attributes of capsules. A capsule may have an internal structure that can be specified using an object diagram or collaboration diagram. The nodes of internal structure are also capsules, which may have an internal structure of their own and so on. This hierarchical decomposition allows the modeling of complex system structures. Fig. 2 shows the structure models of the automatic gauge control system consisting of several active object and interconnection between objects through ports. In UML-RT, The sequence diagram represents the behavior of a capsule. It shows the sequence of messages between objects. The graphical syntax for sequence diagram in automatic gauge control system is shown in Fig. 3. This figure shows all the elements used in most sequence diagrams. The vertical instance lines represent objects participating in the scenario. The horizontal arrows are messages. Each message line starts at the originator object and ends at the target objects and has a message name on the line, such as the *Thickness_Control* object sends the *Get_Steel_Thickness_h()* message to the *Thickness_Sensor* object. In the sequence diagram, the time flows from the top to bottom and the time axis only shows sequence.

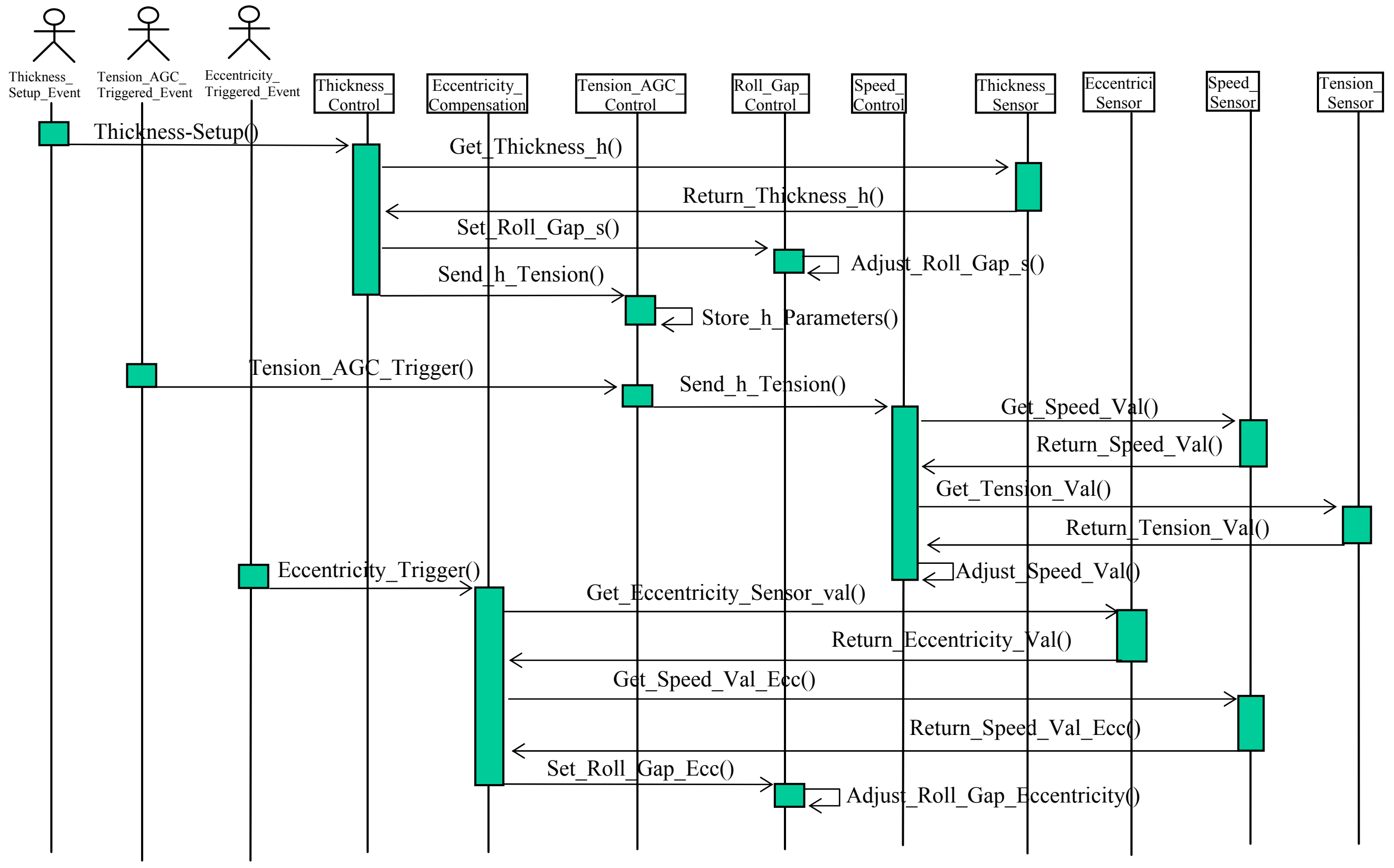

Fig. 3: Sequence Diagram of Automatic Gauge Control System

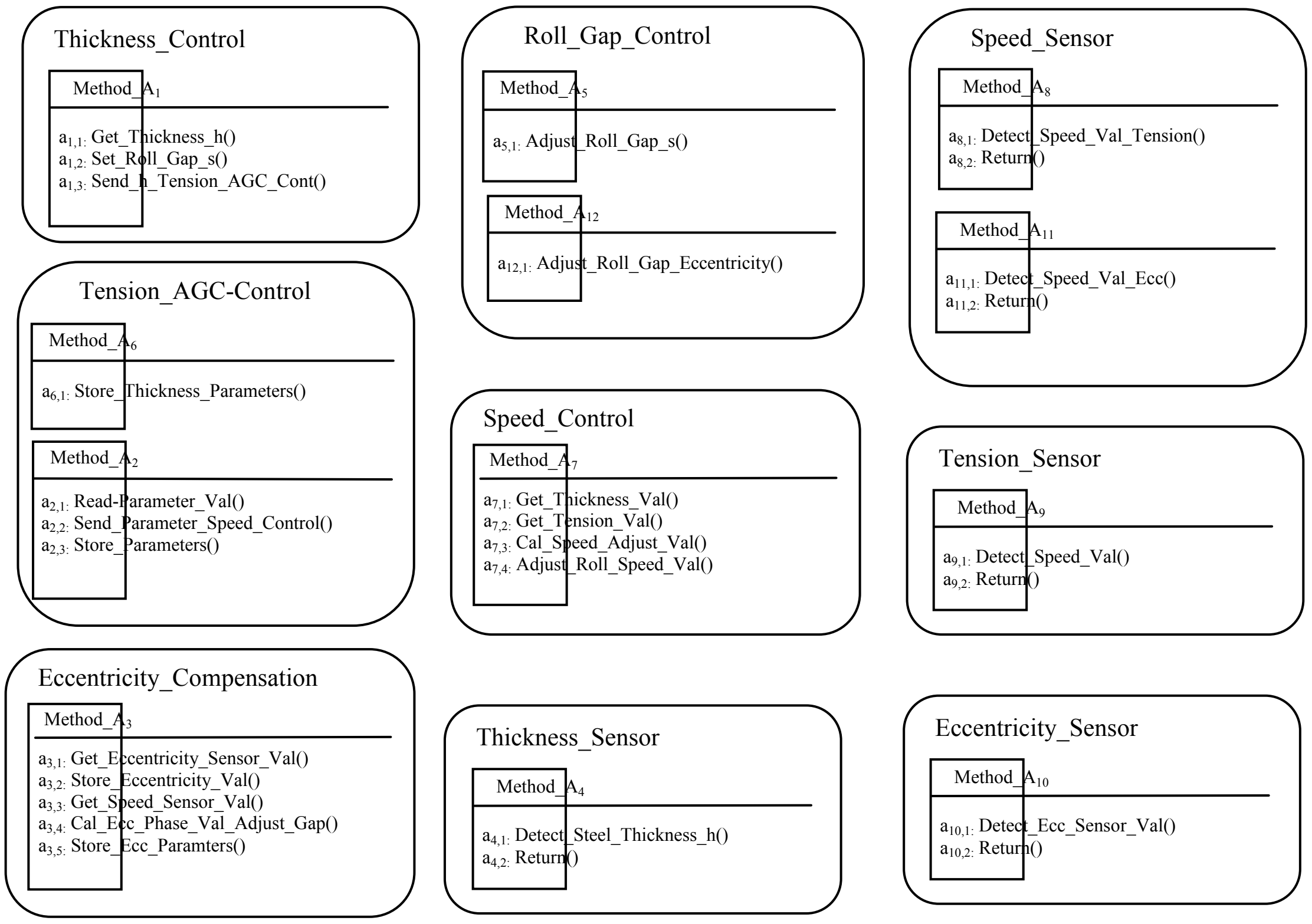

Fig. 4: Method Description of Automatic Gauge Control System

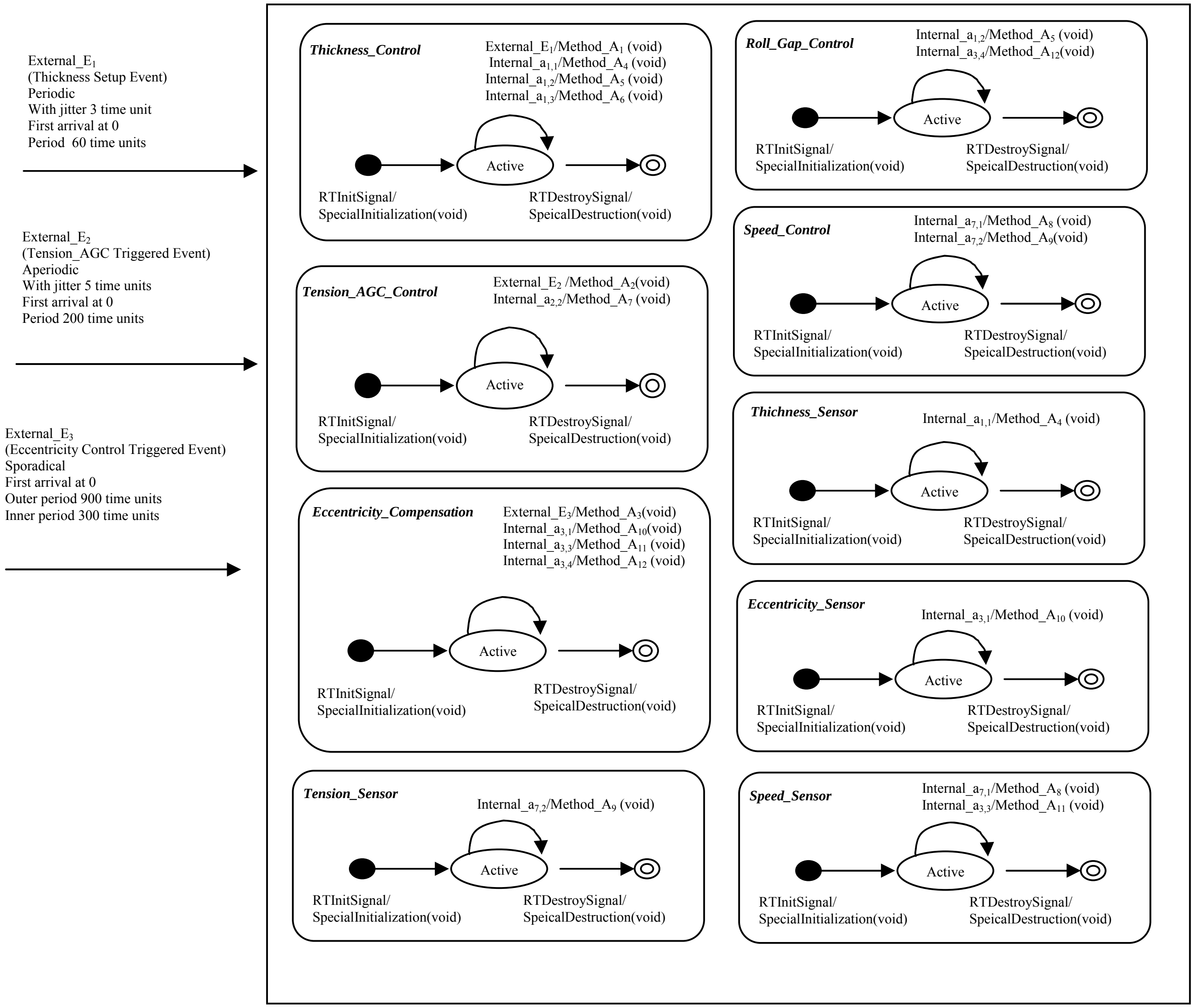

Fig. 5: The General Description of Automatic Gauge Control System

Table 1: Time Characteristics of Automatic Gauge Control System

| Trans | Out.P. | Inn.P. | Num. | Jitter | Event(Type) | Action | Priority | Deadline | Sub-action | Comp.Time | Events Generated |
|---|---|---|---|---|---|---|---|---|---|---|---|
| $\tau_i$ | $T_i$ | $t_i$ | $n_i$ | $J_i$ | $E_i$ | $A_i$ | $\pi(A_i)$ | $D(A_i)$ | $a_{i,j}$ | $C_{i,j}$ | $E_i(a_{i,j})$ |
| $\tau_1$ | 60 | 60 | 1 | 3 | $E_1$ External) | $A_1$ | 10 | 60 | $\{a_{1,1}, a_{1,2}, a_{1,3}\}$ | {5, 1, 1} | $E_4(a_{1,1})$, $E_5(a_{1,2})$, $E_6(a_{1,3})$, |
| | | | | | $E_4$ (call) | $A_4$ | 10 | 60 | $\{a_{4,1}, a_{4,2}\}$ | {5, 1} | --- |
| | | | | | $E_5$ (Signal) | $A_5$ | 10 | 60 | $\{a_{5,1,}\}$ | {5} | --- |
| | | | | | $E_6$ (Call) | $A_6$ | 10 | 60 | $\{a_{6,1}\}$ | {3} | --- |
| $\tau_2$ | 200 | 200 | 1 | 5 | $E_3$ External) | $A_2$ | 9 | 125 | $\{a_{2,1}, a_{2,2}, a_{2,3}\}$ | {4,1,5} | $E_7(a_{2,2})$ |
| | | | | | $E_7$ (Signal) | $A_7$ | 9 | 125 | $\{a_{7,1}, a_{7,2}, a_{7,3}, a_{7,4}\}$ | {4,1,5,1} | $E_8(a_{7,1})$, $E_9(a_{7,2})$ |
| | | | | | $E_8$ ( Call) | $A_8$ | 9 | 125 | $\{a_{8,1}, a_{8,2}\}$ | {6, 1} | --- |
| | | | | | $E_9$ (Call) | $A_9$ | 9 | 125 | $\{a_{9,1}, a_{9,2}\}$ | {8,1} | --- |
| $\tau_3$ | 900 | 300 | 3 | | $E_3$(External) | $A_3$ | 8 | 250 | $\{a_{3,1}, a_{3,2}, a_{3,3}, a_{3,4}, a_{2,5}\}$ | {1,3,1,1,4} | $E_{10}(a_{3,1})$, $E_{11}(a_{3,3})$, $E_{12}(a_{3,4})$ |
| | | | | | $E_{10}$ (Call) | $A_{10}$ | 8 | 250 | $\{a_{10,1}, a_{10,2}\}$ | {7,1} | --- |
| | | | | | $E_{11}$ (Call) | $A_{11}$ | 8 | 250 | $\{a_{11,1}\ a_{11,2}\}$ | {6, 1} | --- |
| | | | | | $E_{12}$ (Signal) | $A_{12}$ | 7 | 250 | $\{a_{12,1}\}$ | {30} | --- |

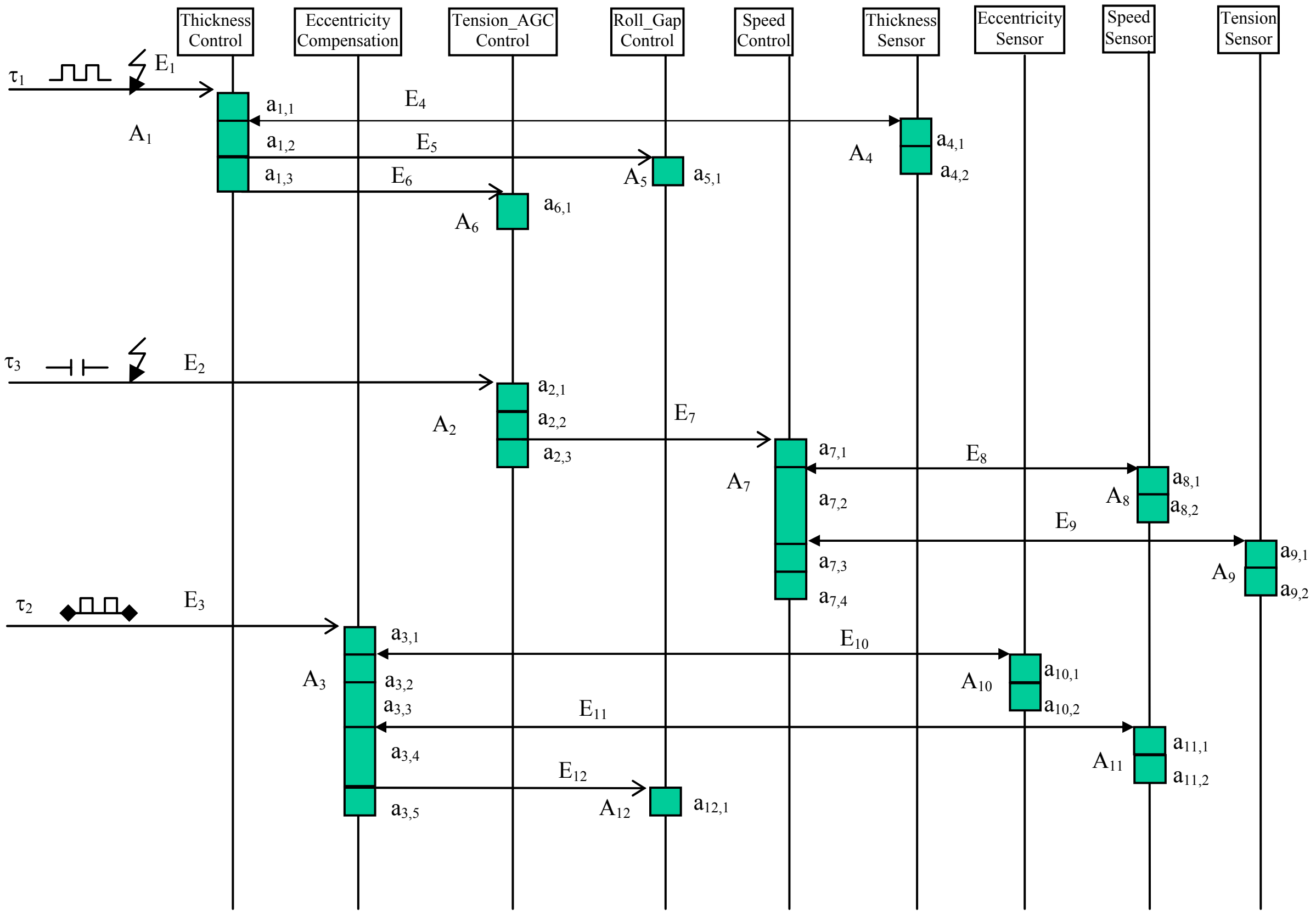

Fig. 6: Extended Sequence Diagram of Automatic Gauge Control System

**Extended Analysis Model for Real-time Control Systems:** Based UML-RT and automatic code generation methodologies and tools, we can automatically produce a feasible real-time control system and executable codes. But for the real-time control systems with release jitter and sporadic effects, we must improve these methodologies, especially in schedulability analysis models and schedulability analysis methods. We developed a schedulability analysis model for real-time control systems based on UML-RT and automatic code generation methodologies. Our schedulability analysis model is restricted to uni-processor single thread implementation environment and it is applicable to the design models and implementation models presented in UML-RT. To facilitate schedulability analysis, our schedulability analysis model can be systematically derived from the application models and implementation models. The analysis gives a view of the real-time control system that focuses on end-to-end behaviors, instead of object behaviors. This is useful since timing constraints in the real-time control systems are often "end-to-end" in nature, i.e., from the system inputs to system outputs and thus, span a computation that may involve the collaboration of multiple objects.

In our study, we assume that real-time control systems are implemented in a uni-processor single thread environment and it is made up of a set of transactions, where transaction denotes a single end-to-end computation within the system. Specifically, it refers to the entire causal set of actions executed as a result of the arrival of an external event that originated from an external source. External event sources are typically input devices (such as sensors) that interrupt the CPU-running embedded software. These external events can be periodic or aperiodic and also have jitter and sporadically periodic characteristics. We express the real-time control system as a collection of transactions that capture all computation in the design model. We also use the term action to capture the processing information associated with an external or internal event. In our model, an action captures this entire run-to-completion processing of an event. The execution of an action may generate internal events that trigger the execution of other actions. Thus, each transaction can be expressed as a collection of actions and events. Each action is a composite action and composed from primitive sub-actions, these primitive sub-actions include send, call and return actions [8], which generate internal events through sending messages to other objects.

**Notation:** In our study, we use event and message as synonymous. Let $\varepsilon = \{E_1, E_2, \ldots, E_n, E_{n+1}, \ldots, E_N\}$ represent the set of all event-streams in the system, where $E_1$, $E_2$, …, $E_n$ denote external event streams and the remaining are internal ones. All external events are assumed to be asynchronous, periodic, aperiodic events and sporadic events with release jitter. We use $J_i$ to represent the jitter time of external event $E_i$. $T_i$ and $t_i$ represents the outer period and inner period for sporadically periodic external events $E_i$. If the external event is without sporadic effects, then inner period of such event is equal to its outer period. Each external event stream $E_i$ corresponds to a transaction $\tau_i$. We also use $A_i$ to represent an action that is associated with each event $E_i$. An action may be decomposed into a sequence of sub-actions $A_i = \{a_{i,1}, a_{i,2}, a_{i,3}, \ldots, a_{i,n_i}\}$, where each $a_{i,j}$ denotes a primitive action, such as sending a message, calling a message and returning a message. We use *q* to represent the instance '*q*' of action $A_i$. Within this model, each action $A_i$ represents the entire "run-to-completion" processing associated with an event $E_i$ and it is characterized as either asynchronously triggered or synchronously triggered, depending on whether the triggering event is asynchronous or synchronous. Each action $A_i$ executes within the context of an active object (capsule) $\tilde{O}(A_i)$ and it is also characterized by a priority ($\pi(A_i)$), which is the same as the priority of its triggering event $E_i$. Each action $A_i$ is also characterized by the computation time ($C(A_i)$) and the deadline ($D(A_i)$). Each sub-action $a_{i,j}$ of $A_i$ is characterized by a computation time $C(a_{i,j})$ (abbreviated as $C_{i,j}$); the computation time of an action is simply the sum of its component sub-actions, i.e.,

$$C(A_i) = \sum_j C_{i,j}$$

also, the computation time of any sequential sub-group of sub-actions $a_{i,p}$ to $a_{i,q}$ where $p \le q$ is:

$$C_{i,p\ldots q} = \sum_{j=p}^{j\le q} C_{i,j} \, .$$

Each event and action is part of a transaction. For the rest of this study, we use superscript to denote transactions. For example, $A_i^\tau$ represents an action and $E_i^\tau$ represents an event, both of which belong to transaction τ. Adding the superscript for external events $\{E_k : k=1, 2, \ldots, n\}$ is unnecessary since there is exactly one external event associated with each transaction, i.e., external event $E_k$ belongs to transaction k and would be denoted as $E_k^k$. In this case, the superscript will be omitted.

**Communication Relationships:** We assumed that there are two types of communication relationships between actions, asynchronous and synchronous. We use symbol "→" to denote asynchronous relationship. An asynchronous relationship $A_i \rightarrow A_j$ indicates that action $A_i$ generates an asynchronous signal event $E_j$ (using a send sub-action) that triggers the execution of action $A_j$. Likewise, we use symbol "↔" to denote synchronous relationship. A synchronous relationship $A_i \leftrightarrow A_k$ indicates that action $A_i$ generates a synchronous call event $E_k$ (using a call sub-action) that triggers the execution of action $A_k$. We assume that if the events have a synchronous relationship, the actions have the same priority. We also use a "causes" relationship and use the symbol ∝ for that purpose. Both asynchronous and synchronous relationships are also causes relationships, i.e., $A_i \rightarrow A_j \Rightarrow (A_i \propto A_j)$ and $A_i \Leftrightarrow A_j \Rightarrow (A_i \propto A_j)$, Moreover, the causes relationship is transitive, thus $(A_i \propto A_j) \wedge (A_j \propto A_k) \Rightarrow A_i \propto A_k$. When $A_i \propto A_j$. We say that $A_j$ is a successor of $A_i$ since $A_i$ must execute (at least partially) for $A_j$ to be triggered.

**Synchronous Set:** For the purpose of analysis, we define the term "synchronous set of $A_i$". The synchronous set of $A_i$ is a set of actions that can be built starting from action $A_i$ and adding all actions that are called synchronously from it. The process is repeated recursively until no more actions can be added to the list. We use $\Upsilon(A_i)$ to denote the synchronous set of $A_i$ and $C(\Upsilon(A_i))$ to denote the cumulative execution time of all the actions in this synchronous set. We also call $A_i$ as the root action of this synchronous set.

**Release Jitter:** The release jitter time is the maximum time that a message may be delayed between the arrival of the message that awakes the transaction and the time the message is put in the run-queue (release). In real-time control systems, the external messages or events may suffer release jitter due to being dispatched by tick driven scheduler. The external events arrivals are not perfect periodic and aperiodic. In our analysis model, we assume that only the external events has release jitter problem and the internal events do not have jitter problem, because the internal event arrival is only decided by the action that represents the entire "run-to-completion" processing associated with the internal event.

**Sporadically Periodic Event:** Sporadically periodic event is a message that arrives at some time and executes periodically for a bounded number of periods (called inner periods) and then re-arrives periodically for a number of times and then does not re-arrive for a larger time (called outer periods). Some real-time control systems have messages that behave as so-called Sporadically periodic messages, example of such messages are interrupt handlers for burst interrupts, packet arrivals from a communication device, or some certain monitoring tasks.

**Extended UML-RT for Control System Modeling:** We know that there are a lot of advantages of UML, UML-RT and automatic code generation methodologies for real-time control system development, such as consistency of model views, problem abstraction, improvement of problem abstract, stability and reusability, automatic code generation. Although these methodologies have a lot of advantages for real-time control systems, the explicit timing requirements in real-time system are not graphically expressed and the release jitter and sporadic effects in real-time control system are not addressed. In this chapter, we use the automatic gauge control system to illustrate our extensions of UML for real-time control systems. This real-time control system has release jitter and sporadic effects.

**General Description of Automatic Gauge Control System:** Figure 4 and 5 give the general description of the automatic gauge control system. This system is made up of nine objects, where each object's finite state machine is shown. We can observe that each objects has only one "real" state associated with it. We also notice that each object calls its *SpecialInitization* action during initialization, through the system event *RTInitSignal* and *SpecialDestruction* action during system shutdown, through the system event *RTDestroySignal.* In addition, there are three external events interacting with the system just described above. The first external is thickness setup event. This event is a periodic event with period 60 time unit and 3 time unit release jitter in the system. The second external event is *Tension_AGC* triggered event, which is an aperiodic event with period 200 time units and 5 time unit release jitter. The third external is Eccentricity Control triggered event; this event is a sporadical event, with outer period 900 time units and inner period 300 time units. The entire external events arrive into the system at time 0.

**Timing Characteristics of Automatic Gauge Control System:** We have described the automatic gauge control system functional requirements. Now, we will consider the timing characteristics of the system, Table 1 shows the timing characteristics in the automatic gauge control system. All the timing properties can be derived from the real-time control system timing requirements. From the table we can see that events have unique priorities, can arrive at any time, but have variable bounded delay before being placed in a priority-order run-queue. Periodic and aperiodic events are given worst-case inter-arrival time and sporadically periodic events are given the outer period and inner period. Each event cannot re-arrive sooner than its inner-arrival time, each event may execute a bounded amount of computation and it is associated with the action, each action is given the worst-case execution time and deadline. This worst-case execution time value is deemed to contain the overhead due to context switching. The cost of pre-emption, within the model, is thus assumed to be zero.

**Extensions of UML-RT for Real-time Control Systems:** From UML and UML-RT, we know that the finite state machine behavior models of objects are useful for code-generation; they are not very conducive for reasoning about end-to-end behaviors, or scenarios. UML-RT uses sequence diagrams to model end-to-end system behaviors, or scenarios. However, sequence diagrams are weak in expressing a detailed specification of end-to-end behaviors, which is necessary for schedulability analysis. To express our ideas, we extend the sequence diagram notation to capture detailed end-to-end behaviors.

In the extended sequence diagram of UML-RT, we capture the details of the processing associated with an event. In the rest of the thesis, we will use the term *action* to refer to the entire run-to-completion processing for an event. Each action is, in general, a composite action and composed from primitive *sub-action*s. These primitive sub-actions include the send, call and return actions described above, which generate internal events through sending messages to other objects. For the purpose of this thesis, we will restrict our attention to a single sequence of sub-actions, although we note that conditional behavior (as may happen with guard conditions associated with transitions) can easily be incorporated. We will also assume that if an action is triggered by a synchronous message (generated from a call action), then that action must have a single reply action and that this action must be the last sub-action. In the extended sequence diagram of UML-RT, we also use the follows notations to represent the different event types.

* We use " → " to represent the asynchronous messages (events).
* We use " ↔ " to represent the synchronous messages (events).
* We use "  " to represent the periodic messages (events).
* We use " ⊣⊢ " to represent the aperiodic messages (events).
* We use " 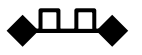 " to represent the sporadically periodic messages (events).
* We use " 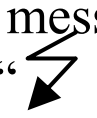 " to represent the release jitter time of messages (events).

Figure 6 describes the automatic gauge control system for rolling mill as discussed. The transaction in the system is driven by different external events. As it can be seen, the *Thickness_Control* object obtains the steel strip thickness from the *Thickness_Sensor* object using a synchronous call action. It then does the control law calculations and generates a roll gap value, which is sent asynchronously to the *Roll_Gap_Control* object, the *Roll_Gap_Control* object is responsible to adjust the gap of roll in the No.1 Stand, then using this method to adjust the thickness of steel strip.

The extended sequence diagram can easily be extended to include sub-actions associated with code executed by the real-time execution framework. In this extended sequence diagram, we can see the external events, internal event, actions and sub-actions. We can also to express the externals event arrival pattern, such as periodic external event with release jitter, aperiodic event with release jitter, sporadic external event with outer period and inner period. The extended sequence diagram is useful to capture timing constraints. Such as arrival rates of external events; periodic, aperiodic and sporadically periodic external messages (events); release jitter time of external messages (events); and end-to-end deadlines. This extended sequence diagram can be integrated with our proposed real-time scheduling algorithms [19] to analysis the schedulability and feasibility of control systems. This extended UML-RT can also be integrated with automatic code generation methodologies to produce code for the feasible control systems. Using this extended UML-RT, designers can quickly evaluate the impact of various implementation decisions on schedulability. In conjunction with automatic code-generation, we believe that this will greatly streamline the design and development of real-time control system software.

## CONCLUSION

Software design has become more and more important within the real-time control system design process since functionality implementation gradually migrated from hardware to software. Consequently, several commercial tools have become available that provide an integrated development environment for real-time control systems with object-oriented techniques to facilitate the design phase, e. g., Artisan Real-Time Studio (http://www.artisansw.com) and Rational Rose Real-Time (http://ibm.com). However, these tools lack the 'real-time" support required by many of these systems, especially those with stringent timing constraints.

This work put forward a formal model for specifying timeliness properties. The application of the model is shown within a realistic case study. We have extended UML sequence diagrams to visually describe the timing properties for real-time control systems. The proposed notation is generally applicable to any modeling language using active objects and explicit communication between objects through message passing. This method can be used to cope with timing constraints in realistic and complex real-time control systems.